%% file: Main.tex
\documentclass[12pt, oneside]{report}

\usepackage[utf8]{inputenc}
\usepackage{comment}
\usepackage{nameref}
\usepackage{graphicx}
\usepackage{mathtools}
\usepackage{dsfont}
\usepackage{fancyhdr}
\usepackage{lipsum}
\usepackage{hyperref} 
\usepackage{titletoc}
\usepackage{tcolorbox} 
\usepackage{natbib}
\usepackage{enumitem}
\usepackage{geometry}

\usepackage{array}    
\usepackage{dsfont}
\usepackage{booktabs}
\usepackage{makecell} 
\usepackage{float}
\usepackage{titletoc}
\usepackage{hyperref}
\usepackage{cleveref}
\usepackage[toc,page]{appendix}
\usepackage{bbm}
\usepackage{amsfonts}
\usepackage{stackengine}
\usepackage{amsmath}
\usepackage{amsthm}
\usepackage{amssymb}
\usepackage{colortbl}
\usepackage{xcolor}
\usepackage{tcolorbox} % For the box around contents
\usepackage{subcaption} % for subfigures
\usepackage{caption} 
\theoremstyle{definition}

\newcommand{\bfcell}[1]{\textbf{#1}}

\usepackage{titlesec}
\titleformat{\chapter}[display]
  {\normalfont\LARGE\bfseries}{Chapter \thechapter}{15pt}{\LARGE}
\titlespacing*{\chapter}{0pt}{40pt}{30pt}

\arrayrulecolor{black}

\renewcommand{\thesection}{\thechapter.\arabic{section}}
\renewcommand{\thesubsection}{\thesection.\arabic{subsection}}

\newcommand{\rowspace}{\rule{0pt}{18pt}}

\titleformat{\section}{\Large\bfseries}{\thesection}{1em}{}

\titleformat{\subsection}{\Large\bfseries}{\thesubsection}{1em}{}

\titleformat{\subsubsection}{\large\bfseries}{\thesubsubsection}{1em}{}

\titleformat{\paragraph}[runin]{\normalfont\normalsize\bfseries}{\theparagraph}{1em}{} % Formatting for \paragraph
\titlespacing*{\paragraph}{0pt}{3.25ex plus 1ex minus .2ex}{1em} % Adjust spacing for \paragraph

\newcommand{\applycolor}[1]{%
    \ifdim #1 pt > 0.8 pt
        \cellcolor{red!70}#1%
    \else\ifdim #1 pt > 0.6 pt
        \cellcolor{red!50}#1%
    \else\ifdim #1 pt > 0.4 pt
        \cellcolor{orange!50}#1%
    \else\ifdim #1 pt > 0.2 pt
        \cellcolor{orange!40}#1%  % Slightly darker orange
    \else\ifdim #1 pt > 0.15 pt
        \cellcolor{orange!35}#1%  % Slightly darker orange
    \else\ifdim #1 pt > 0.09 pt
        \cellcolor{orange!25}#1%  % Light orange
    \else
        \cellcolor{yellow!30}#1%
    \fi\fi\fi\fi\fi\fi
}

\arrayrulecolor{black}

\titleformat{\chapter}[display]
  {\normalfont\LARGE\bfseries}{}{0pt}{\LARGE}

\renewcommand{\thesection}{\arabic{section}}
\renewcommand{\cdot}{}

\begin{document}

%\pagenumbering{gobble}

%\maketitle

{\Large 
A Bayesian Modelling Framework 
with Model Comparison for Epidemics with Super-Spreading}

\vspace*{2cm}

Hannah Craddock$^1$, Simon E.F.~Spencer$^1$, Xavier Didelot$^{1,2,*}$

\vspace*{2cm}

$^1$ Department of Statistics, University of Warwick, United Kingdom\\
$^2$ School of Life Sciences, University of Warwick, United Kingdom\\
$^*$ Corresponding author. Tel: 0044 (0)2476 572827. \\
Email: \verb+xavier.didelot@warwick.ac.uk+

\vspace*{2cm}
Running title: Bayesian modelling of super-spreading epidemics

\vspace*{1cm}

\newpage

\section*{ABSTRACT}
\input{ABSTRACT}

%\tableofcontents 

\newpage

\section*{INTRODUCTION}
\input{1_INTRODUCTION}

\section*{MATERIAL AND METHODS}
\subsection*{Modelling Framework Overview}
\input{2_MODELS}

\subsection*{Bayesian Inference Methodology}
\input{3_INFERENCE}

\subsection*{Model Comparison}
\input{4_MODEL_COMPARISON}

\subsection*{Implementation}
\vspace*{-0.5cm}
We implemented the analytical methods described in this paper in a 
new R package which is available
at \url{https://github.com/hanmacrad2/SuperSpreadingEpidemicsMCMC} for R version 3.5 or later. 
All code and data needed to replicate the results are included in this repository.

\section*{RESULTS}
\input{5_REAL_DATA_APPLICATIONS}

\section*{DISCUSSION}
\input{6_DISCUSSION}

\bibliography{References}

\end{document}

%% file: ABSTRACT.tex
The transmission dynamics of an epidemic are rarely homogeneous. Super-spreading events and super-spreading individuals are two types of heterogeneous transmissibility. Inference of super-spreading is commonly carried out on secondary case data, the expected distribution of which is known as the offspring distribution. However, this data is seldom available. Here we introduce a multi-model framework fit to incidence time-series, data that is much more readily available. The framework consists of five discrete-time, stochastic, branching-process models of epidemics spread through a susceptible population. The framework includes a baseline model of homogeneous transmission, a unimodal and a bimodal model for super-spreading events, as well as a unimodal and a bimodal model for super-spreading individuals. Bayesian statistics is used to infer model parameters using Markov Chain Monte-Carlo. Model comparison is conducted by computing Bayes factors, with importance sampling used to estimate the marginal likelihood of each model. This estimator is selected for its consistency and lower variance compared to alternatives. Application to simulated data from each model identifies the correct model for the majority of simulations and accurately infers the true parameters, such as the basic reproduction number. We also apply our methods to incidence data from the 2003 SARS outbreak and the Covid-19 pandemic caused by SARS-CoV-2. Model selection consistently identifies the same model and mechanism for a given disease, even when using different time series. Our estimates are consistent with previous studies based on secondary case data. Quantifying the contribution of super-spreading to disease transmission has important implications for infectious disease management and control. Our modelling framework is disease-agnostic and implemented as an R package, with potential to be a valuable tool for public health.

%% file: 1_INTRODUCTION.tex
When an epidemic outbreak occurs, the transmission dynamics are rarely homogeneous. During the Covid-19 pandemic of SARS-CoV-2 for example, it became evident that super-spreading events played a crucial role in the outbreak and early on signs of super-spreading were reported \citep{endo2020estimating,wang2020inference,du2022systematic}. Events such as weddings, family gatherings and sports events, in which many people are infected at once, spawned dangerous outbreaks \citep{lewis2021superspreading}. Such uneven transmission dynamics are common among the Coronavirus's relatives, including SARS-CoV-1, responsible for the severe acute respiratory syndrome (SARS) epidemic in 2003, and MERS-CoV, which causes the Middle East respiratory syndrome \citep{WANG20215039,brainard2023super}. 

A key parameter in understanding the transmission dynamics is the basic reproduction number $R_0$, the average number of secondary infections caused by one infected individual in a population where all individuals are susceptible to infection. An estimate of $R_0$ can help establish if there is a high probability of a major outbreak occurring and can provide feedback on the success of control interventions given that the goal of such efforts is to reduce $R_0$ below the threshold value of $1$ \citep{fraser2004factors,ferguson2006strategies,fraser2009pandemic}.
The most common modelling approach in epidemics is to use a transmission or compartmental model \citep{keeling2011modeling}. However, branching process models are a flexible and biologically-realistic alternative \citep{farrington2003branching}. They are particularly suited for modelling the early stages of an outbreak, for studying $R_0$ and transmission heterogeneity which is why we used them in this research. For example the widely used epidemic modelling tool developed by \citet{cori2013new} uses a branching process model to estimate $R_t$, the time-varying reproduction number over the course of an epidemic. The framework is disease agnostic and uses as input case incidence data, as we do here, as opposed to contact tracing or secondary case data, which is less readily available. 

The modelling of super-spreading in epidemics is present in the literature but remains relatively limited \citep{grassly2008mathematical}. The work of \citet{lloyd2005superspreading} was a seminal paper on this topic. The authors compared the fit of three different offspring distributions -- Poisson, geometric and negative binomial -- to data on the number of secondary cases. However this approach requires access to contact tracing data, which is not often readily available. Furthermore the negative binomial distribution, despite being the best fitting model, fails to capture possible bimodality of the offspring distribution. More recently, the work of \citet{ho2023accounting} on super-spreading and over-dispersion also applied the negative binomial, but directly to incidence data to estimate $R_t$. However, no comparison is made with competing alternate models. Here we perform Bayesian inference and comparison between five models: a baseline model without any transmission heterogeneity, two models of super-spreading individuals (unimodal and bimodal) and two models of super-spreading events (unimodal and bimodal). We apply this framework to several simulated and real datasets to showcase its usefulness. Given the documented prominence of super-spreading in the recent epidemics of the 21st century, quantifying the contribution of super-spreading on disease transmission has important implications for the control and management of infectious diseases. 

%% file: 2_MODELS.tex
%MODELS
\input{SECTION_MODELS/I_INTRODUCTION_MODELS}

\input{SECTION_MODELS/II_INFECTIOUSNESS}

%%%%%%%%%%%%%%%%%%%%
% SUMMARY OF MODELS

\begin{figure}[!p]
\centering
\includegraphics[height=0.72\textwidth]{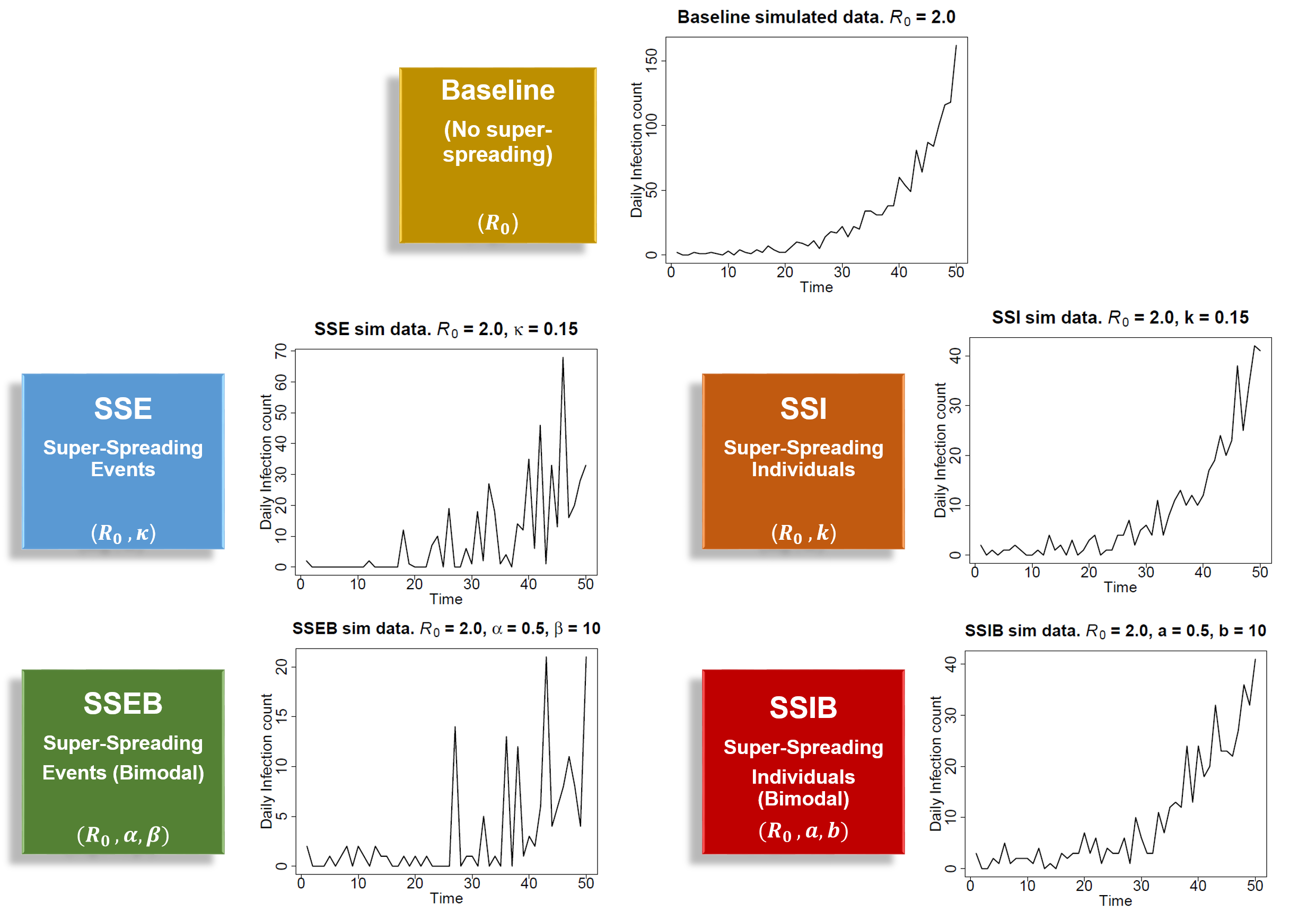}
\vspace{2mm} 
\caption{Exemplary simulation from the five models of epidemic transmission. The time series plots show incidence data $\boldsymbol{I}_{[1:T]}$ simulated from the Baseline, SSE, SSI, SSEB and SSIB models for $R_0$ = $2.0$, duration $T = 50$.   %We select suitable parameter values for the simulations, guided by \textbf{Table Supp 1} %\autoref{tab:table_epi_params} 
%for the parameters $R_0$ and $k$ and carefully considered values for the novel parameters.
}
\label{fig:model_framework}
\end{figure}

\input{TABLES/table_models_summary}

\subsection*{The Baseline Model}\input{SECTION_MODELS/1_BASELINE_MODEL}

\subsection*{The SSE Model}\input{SECTION_MODELS/2_SSE_MODEL}

\subsection*{The SSI Model}\input{SECTION_MODELS/3_SSI_MODEL}

\subsection*{The SSEB Model}\input{SECTION_MODELS/4_SSEB_MODEL}

\subsection*{The SSIB Model}\input{SECTION_MODELS/5_SSIB_MODEL}

%% file: SECTION_MODELS/I_INTRODUCTION_MODELS.tex
We present a framework of five discrete-time, stochastic branching process models to describe infectious disease transmission through a susceptible population. The epidemiological process considered is a branching process whereby each infected individual transmits to secondary cases called offspring \citep{farrington2003branching}. The framework is adaptable to a range of infectious diseases. %A gap in the literature was identified relating to models of super-spreading transmission, particularly relating to two key points;
%
%\textbf{Data}: 
Our models are fit to incidence data, which represents reported cases over time (e.g., daily cases). %Secondary case data, derived from contact tracing, is less available, resource-intensive to collect and is often uncertain—identifying who infected whom is challenging. Only 8 countries were reported to have collected detailed contact-tracing data for SARS-CoV-2 \citep{du2022systematic}. The offspring distribution $Z$, describing the number of secondary infections per individual, whereby \( \mathbb{E}(Z) = R_0 \), is important for understanding epidemic dynamics but is not used for model fitting here. Incidence data is more widely recorded, making it our preferred choice. 
We define the incidence data \( I_t \) as the number of infections at time \( t \) and the total incidence data up to time \( T \) as \( \boldsymbol{I}_{[1:T]} = [I_1, I_2, \dots, I_T] \).  

%\textbf{Models}: 
To account for super-spreading or over-dispersion in transmission, the negative binomial model is often chosen to model the offspring distribution $Z$, describing the number of secondary infections per individual \citep{lloyd2005superspreading, endo2020estimating}. However, this model is uni-modal and so it is unable to generate multiple or even secondary modes to account for additional super-spreading phenomena. We seek to address this limitation by introducing two novel bimodal models of epidemic transmission.
Our model framework of epidemic transmission consists of a Baseline model with no super-spreading properties, two models that describe super-spreading events (SSE and SSEB) and two models that describe super-spreading individuals or infections (SSI and SSIB). We define a super-spreading event (SSE) as an event or point in time in which a large number of infections are generated. Such events could include weddings, family gatherings and sports events, in which many people were infected at once \citep{lewis2021superspreading}. A super-spreading individual (SSI) refers to an individual who is more infectious than non super-spreading individuals in the population. \citet{wallinga2004different} defines such a super-spreading individual as one that produces at least 10 secondary infections. The models also vary by mode; the Baseline, SSE, and SSI models are uni-modal, using the Poisson and negative binomial distributions. The bimodal (suffix `B') models, the SSEB and SSIB models, introduced in this work, offer novel approaches for modelling super-spreading events and individuals. Exemplary simulations from each of the five models are displayed in \autoref{fig:model_framework}. 
For the super-spreading events models (SSE and SSEB), the spikes in infections at certain time-points corresponding to super-spreading events are evident. 
For the SSI and SSIB models, the increased infectivity of super-spreading individuals lasts the duration of their infectious period, so spikes in infections do not occur at a specific time point, 
but rather are spread across the duration of their infectivity.
The models are detailed below in turn and summarized in \autoref{tab:table_models}. %Full model derivations can be found in the supplementary material.

%% file: SECTION_MODELS/II_INFECTIOUSNESS.tex
Generating incidence data $I_t$ at time $t$ from each model requires calculating the infectious pressure from individuals infected at earlier time points. 
%We first outline the infectious contribution of a single individual over their infectious period and then extend this to the entire population of infected individuals over the course of the epidemic. 
We assume that each infected individual has an infectivity profile given by a probability distribution denoted $\omega({\tau})$, dependent on the time since infection time $\tau$,
as in many previous studies \citep{wallinga2004different,cori2013new,didelot2017genomic}. %An individual will be most infectious at time $\tau$ when $\omega({\tau})$ is at its maximum. A similar approach is seen in \citet{}. 
%To accommodate discrete-time incidence data, we adopt a discretized gamma function, parameterized by shape $\theta$ and scale $\sigma$ as follows:
%\begin{equation}
%\omega(\tau) = F_{\mathrm{Gamma}}(\tau; \theta, \sigma) - F_{\mathrm{Gamma}}(\tau - 1; \theta, \sigma) \hspace{3mm} \label{eq:omega_discrete}
%\end{equation}
%
%This formulation ensures that $\sum_{\tau = 1}^{\infty} \omega(\tau) = 1$, representing the complete infectious lifetime. In this study, we use a Gamma distribution with $\theta = 6$, $\sigma = 1$, yielding a mean of 6, a mode of 5, and a standard deviation of 2.45. Such summary statistics align with serial interval estimates for diseases like SARS-CoV-2 and SARS (see Figure X in Supplement Material).
%
%Population Level Infectiousness across Time: 
The total infectious pressure at time $t$, $\lambda_t$, is the cumulative contribution of individuals infected at earlier time points. Each infected individual contributes $\omega(t - \tau)$ to $\lambda_t$, defined for $t = 2, \dots, T$. We assume no infectious pressure from individuals infected prior to time $t = 1$. The total infectious pressure $\lambda_t$ is therefore defined as:
\begin{equation}
    \lambda_t = \displaystyle\sum_{\tau = 1}^{t-1}I_{\tau}\cdot \omega(t-\tau)%, \hspace{5mm} \text{ for  } t = 2 \dots T. \label{eq: lambda}
\end{equation}
\newpage
%As an example consider the total infectious pressure $\lambda_3$ at time $t = 3$ due to past infections; $    \lambda_3 = \displaystyle\sum_{\tau = 1}^{2}I_{\tau}\cdot \omega(3-\tau) = I_1 \cdot \omega(2) + I_2 \cdot \omega(1)$.

%% file: TABLES/table_models_summary.tex
%\newgeometry{top=1.5cm, bottom=2cm, left=1.5cm, right=1.5cm}
\newcommand{\custombig}{}%\fontsize{13}{15}\selectfont}

\begin{table}[!p]
\centering
\hspace*{-2cm}\begin{tabular}{|p{43mm}|p{42mm}|p{23mm}|p{69mm}|}
\hline
\rowspace \newline \custombig \centering\arraybackslash\bfcell{Model} \newline & \rowspace \newline \custombig \custombig \centering\arraybackslash\bfcell{Parameters} \newline & \rowspace  \newline \custombig \centering\arraybackslash\bfcell{Parameter Range of}
\newline \centering\arraybackslash\bfcell{  Interest} \newline & \rowspace \newline \custombig \centering\arraybackslash\bfcell{Incidence Data} \newline \\ 
\hline
%\specialrule{0.15em}{0.2em}{0.2em} % Add horizontal line
\rowspace
\textbf{Baseline} \newline No Super-spreading & ${R_0}$: Basic \newline  Reproduction Number & (0, 10] & $I_t \sim$ Poisson($R_0 \lambda_t$) \\
\hline
\rowspace \textbf{SSE} \newline Super-spreading events \newline & ${R_0}$
\newline 
\rowspace ${k}$: Dispersion\newline Parameter  & (0, 10] \newline \rowspace (0, 1] & $I_t \sim$ NegBin($r=k \lambda_t$, $\mu = R_0 \lambda_t$) \\
\hline
\rowspace   
\textbf{SSI} \newline Super-spreading individuals \newline & ${R_0}$
\newline 
\rowspace ${k}$  & (0, 10] \newline \rowspace (0, 1] &  $I_t | \boldsymbol{\nu^{+}}_{[1:t-1]} \sim \text{Poisson}\big( \displaystyle\sum_{\tau=1}^{t-1} \nu^{+}_{\tau} \cdot \omega(t-\tau)\big) $ \newline \newline $\boldsymbol{\nu^{+}}_{[1:T]} | I_t \sim \mathrm{Gamma}(I_t \cdot k, R_0/k)$ 
\newline \newline $\boldsymbol{\nu^{+}}_t$ : sum of individual reproduction numbers $\nu$ on day $t$ \\
\hline
\rowspace 
\textbf{SSEB} \newline Super-spreading events bimodal \newline \newline & ${R_0}$ \newline \newline ${\alpha}$: Proportion \newline of $R_0$ due to non-SSE infections \newline \newline ${\beta}$: Average size of a SSE
& (0, 10] \newline \newline  [0, 1] \newline \newline  \newline \newline  (1, 20] & $I_t = N_t + S_t$  \newline \newline $N_t \sim \text{Poisson}(\alpha \cdot R_0 \cdot \lambda_t)$ \hspace{2mm} \newline Infections from Non SSEs \newline \newline $S_t \sim \text{Poisson}(\beta \cdot E_t) $ \newline Infections from SSEs  \newline  \newline $E_t \sim \text{Poisson}(R_0(1 - \alpha)/\beta \cdot \lambda_t)$  \newline Number of SSEs\\
\hline 
\rowspace
\textbf{SSIB} \newline Super-spreading individuals bimodal \newline  \newline & ${R_0}$  \newline \newline ${a}$: Proportion \newline of $R_0$ due to non-SSI infections \newline \newline ${b}$: Increased infectivity of SSIs & (0, 10] \newline \newline  [0, 1] \newline \newline  \newline \newline  (1, 20] & $I_t = N_t + S_t$  \newline \newline $N_t \sim \text{Poisson}(a R_0 \lambda_t^{'})$ \newline Non-SSI Infections \newline \newline $S_t \sim \text{Poisson}(R_0(1 - a)/b \cdot \lambda_t^{'}) $ \newline SSI Infections \newline \newline $\lambda_{t}^{'} = \displaystyle\sum_{\tau=1}^{t-1} \bigg( N_{\tau} + b \cdot S_{\tau} \bigg) \cdot \; \omega(t-\tau)
$ \\
\hline
\end{tabular}
%\captionsetup{font=large} % Adjust caption font size
\caption{The five epidemic transmission models fit to incidence data, their parameters and distributions used.}
\label{tab:table_models}
\end{table}

%% file: SECTION_MODELS/1_BASELINE_MODEL.tex
%In the Baseline model, both the offspring distribution \( Z \) and incidence data \( I_t \) follow Poisson distributions. 
The Poisson model is commonly used to capture the stochasticity of epidemic transmission \citep{diekmann2000mathematical, lloyd2005superspreading}, however as its mean equals its variance, it cannot capture transmission heterogeneity. Thus, it serves as our baseline for comparison. The offspring distribution of an individual is:
\begin{equation}
    Z \sim \text{Poisson}(R_0)
    \label{eq:z_baseline}
\end{equation}

The mean of a Poisson distribution is equal to the parameter $R_0$ which is the only parameter of this model.
%\noindent \textbf{The Incidence Model}: 
To generate incidence data $I_t$ from the Baseline model, we use the fact that a sum of Poisson-distributed random variables is also Poisson-distributed, so that:
\begin{equation}
I_t \sim \text{Poisson}(R_0 \cdot \lambda_{t})% \hspace{5mm}
\label{eq:baseline_it}
\end{equation}

%% file: SECTION_MODELS/2_SSE_MODEL.tex
The SSE model is a unimodal model for super-spreading events. %, with both the offspring distribution \( Z \) and incidence data \( I_t \) following negative binomial distributions. 
Unlike the Poisson distribution, the negative binomial distribution has differing mean and variance, allowing for over-dispersion in the number of secondary infections transmitted. A key parameter of this model is the dispersion parameter $k$, a parameter used to quantify heterogeneity in certain distributions \citep{lloyd2007maximum}. In the SSE model the offspring distribution is defined as:
\begin{equation}
Z \sim \text{NegativeBinomial}(r=k, \mu = R_0)
\label{eq:z_sse}
\end{equation}

%Small values of $k$ %($k \rightarrow 0$,  Var $\rightarrow \infty$)
%are indicative of increased heterogeneity in transmission and potential for super-spreading in which a few infectious cases account for the majority of secondary transmissions \citep{du2020serial}. 
%
%
%\noindent \textbf{The Incidence Model}: 
In the SSE model, the incidence data \( I_t \) is modelled as a negative binomial random variable. While this distribution is commonly applied to \( Z \), its use for \( I_t \) is less frequent. A notable example is \citet{ho2023accounting}, where incidence data is employed to estimate \( R_t \), the time-varying reproduction number. In our SSE model, we adopt a similar parameterization involving \( R_0 \), parameterized by size \( r \) and mean \( \mu \), with parameters \( R_0 \) and \( k \):
\begin{equation}
I_t \sim \text{NegativeBinomial} \bigg(r = k \lambda_t, \hspace{2mm} \mu = R_0 \lambda_t \bigg)  
\label{eq:sse_it}
\end{equation}

As \( k \) decreases, the variance increases, signaling greater transmission heterogeneity and potential for super-spreading \citep{du2020serial}. Accurate estimation of \( k \) is crucial for determining the need for public health measures; significant outbreaks may occur when \( k \) is small, even if \( R_0 < 1 \) \citep{kucharski2015role}. 

%% file: SECTION_MODELS/3_SSI_MODEL.tex
The SSI model is a unimodal model for  super-spreading individuals. In this model super-spreading individuals arise from the right-hand tail of the infectivity $\nu$ rather than forming a distinct group. Building on \citet{lloyd2005superspreading}, this model uses the same negative binomial distribution for $Z$ and extends it by introducing incidence data $I_t$ derived from the model. In \citet{lloyd2005superspreading}, $\nu$ is introduced as a random variable representing the expected number of secondary cases from an individual, with $Z \sim \text{Poisson}(\nu)$. Unlike simpler models where $\nu = R_0$, the SSI model assumes $\nu$ is drawn from a Gamma distribution with mean $R_0$ and shape and scale parameters $\alpha$ and $\theta$, respectively: 
\begin{equation}
\nu \sim \mathrm{Gamma}\left(\alpha = k, \theta = \frac{R_0}{k}\right) \label{eq: nu_ssi}
\end{equation}

The offspring distribution $Z$ follows a Poisson distribution with rate $\nu$ which is gamma distributed and is therefore a negative binomial random variable:
\begin{equation}
%Z & \sim \text{Poisson}(\nu) \\
%Z & \sim \text{Poisson}\left(\text{Gamma}\left(k, \frac{R_0}{k}\right)\right) \\
    Z  \sim \text{NegativeBinomial}(r=k,\mu = R_0)
\label{eq:z_ssi}    
\end{equation}

As before, small values of $k$ correspond to high levels of heterogeneity in transmission. $Z$ is the same offspring distribution as in the SSE model, however the models differ in how the incidence data $I_t$ is derived.
%
%\noindent \textbf{The Incidence Model}: 
To generate incidence data $I_t$ from the SSI model, we introduce a new variable $\nu^{+}_t$ as the sum of all individual reproduction numbers $\nu_i$ of each individual $i$ infected at time $t$: $\nu^{+}_t = \sum_{i=1}^{\infty}\nu_i \mathds{1}_{ \{i \text{ infected on day t} \}}$. By applying the scaling properties of the gamma distribution to the distribution of $\nu$ in \autoref{eq: nu_ssi} and accounting for all infected individuals on day $t$, we derive:
\begin{equation}
\nu^{+}_t | I_t \sim \mathrm{Gamma}(\alpha=I_t \cdot k, \theta= R_0/k)
\label{eq:nu_plus_gamma}
\end{equation}

%This approach replaces individual $\nu_i$ values with $\nu^{+}_t$, improving inference speed as we only infer the parameters of $\nu^{+}_t$. 
We record $\nu^{+}_t$ for the duration of the epidemic in the following vector $\boldsymbol{\nu^{+}}_{[1:T]} = [\nu^{+}_1, \nu^{+}_2, \ldots, \nu^{+}_t, \ldots, \nu^{+}_T]$. % \hspace{5mm} \text{ for  } t = 2 \dots T$. 
Incidence data $I_t$ from the SSI model depends on past $\nu^{+}_t$ values. 
%Unlike the Baseline model, where $R_0$ is constant, here 
$I_t$ follows a Poisson distribution with a rate based on the gamma-distributed $\nu^{+}_t$:
\begin{equation}
I_t | \boldsymbol{\nu^{+}}_{[1:t-1]}  \sim \mathrm{Poisson} \bigg( \displaystyle\sum_{\tau=1}^{t-1} \nu^{+}_{\tau} \cdot \omega(t-\tau) \hspace{1.5mm} \bigg)% \hspace{4mm} \text{for } t = 2 \dots T
%\\\nu^{+}_t | I_t & \sim \mathrm{Gamma}(I_t \cdot k, \hspace{2 mm} R_0/k)
\label{ssi_it}
\end{equation}

Once $I_t$ is generated at time $t$, we draw $\nu^{+}_t$ using \autoref{eq:nu_plus_gamma}. The generation of incidence data from the SSI model, assuming known incidence data $I_1$ at time $t = 1$ and model parameters $R_0$ and $k$, is achieved by iteratively sampling from the last two equations.
%is as follows;
%\begin{align}
%\text{For } & t = 2 \text{ to } T: \nonumber \\
%& \text{Draw the sum of the individual reproduction numbers which depends on } I_{t-1}; \nonumber \\ 
%& \nu^{+}_{t-1} \sim \text{Gamma}\left(\theta = I_{[t-1]} \cdot k, \sigma = \frac{R_0}{k}\right) \quad &  \nonumber \\
%& \text{Update the vector } \boldsymbol{\nu^{+}}_{[1:t-1]} ; \nonumber \\
%& \boldsymbol{\nu^{+}}_{[1:t-1]} = [\nu^{+}_1, \nu^{+}_2, \ldots, \nu^{+}_{t-1}]  \nonumber \\
%& \text{Generate the incidence data at time } t \nonumber \\
%& I_t | \boldsymbol{\nu^{+}}_{[1:t-1]} \sim \text{Poisson}\left(\sum_{\tau=1}^{t-1} \boldsymbol{\nu}^{+}_{\tau} \cdot \omega(t - \tau)\right)  \nonumber 
%\end{align}

%% file: SECTION_MODELS/4_SSEB_MODEL.tex
The SSEB model is a bimodal model for super-spreading events. % with three parameters $R_0$, $\alpha$ and $\beta$. We define super-spreading events as events or point in time in which many susceptibles are infected at once. 
In the SSEB model, the total infections at time \( t \), \( I_t \), arise from two mechanisms: homogeneous transmission and super-spreading. These are represented as two independent Poisson processes, \( N_t \) (homogeneous) and \( S_t \) (super-spreading), such that \( I_t = N_t + S_t \). $E_t$ denotes the number of super-spreading events at time \( t \), each of which causes an increased number of infections captured by the parameter \( \beta \). %The model parameters are the reproduction number \(R_0\), \(\alpha\), and \(\beta\).
%\begin{itemize}
The parameter \( \alpha \) is the proportion of \( R_0 \) attributable to homogeneous transmission, with \( 1 - \alpha \) representing the contribution from super-spreading events. %A smaller \( \alpha \) indicates greater influence of SSEs. The Baseline model is a special case of the SSEB model, where \( \alpha = 1 \).
 \( \beta \) reflects the increased number of infections for each super-spreading event. %Spikes in infection counts from SSEs are proportional to \( \beta \), which can be observed in simulations from the model.
%\end{itemize}
%
%\noindent \textbf{The Incidence Model }: 
%As the SSEB model is based on events in time at a population level we start our model derivation from the temporal incidence data followed by the offspring distribution. %The assumption of independence between $N_t$ and $S_t$ arises because infections from homogeneous mixing (\(N_t\)) are not temporally clustered and occur independently of the infections caused by super-spreading events (\(S_t\)). If \(N_t\) were temporally clustered, they would instead be categorized as super-spreading events. \\

%\noindent $N_t$ -- Infections from Non Super-Spreading Events:  

The incidence data from non-SSE events follows a Poisson distribution:
\begin{equation}
N_t \sim \text{Poisson}(\alpha \cdot R_0 \cdot \lambda_{t}) \hspace{5mm}% \text{for} \hspace{2mm} N_t = 0, 1, ..., I_t. 
\label{eq: sseb_nt}
\end{equation}

%\noindent $S_t$ -- Infections from Super-Spreading Events: 

\( S_t \) represents the infections resulting from super-spreading events at time \( t \).  We define $E_t$ as the total number of such super-spreading events at time $t$, for example a concert or a wedding \citep{adam2020clustering}. Each event contributes an increased number of infections, quantified by the parameter \( \beta \). Specifically, each super-spreading event results in a Poisson-distributed number of infections with mean $\beta$. It follows that $S_t$ is a compound Poisson distribution \citep{adelson1966compound}. The contribution to $R_0$ from super-spreading events is $R_0(1 - \alpha)$, and since an SSE yields $\beta$ times more infections:
\begin{equation} 
%\text{Number of super-spreading events: } 
E_t \sim \mathrm{Poisson} \left( \frac{R_0(1 - \alpha)}{\beta} \cdot \lambda_t \right)
\end{equation}
\begin{equation}
%\text{Number of infections by all SSEs at time t: } 
S_t | (E_t=e_t) \sim \mathrm{Poisson}(\beta \cdot e_t)
\label{eq:sseb_st}
\end{equation}

For the purposes of implementation let the maximum possible number of super-spreading events at any given time $t$ be $M$ events, we can then decompose:
\begin{equation}
p(S_t = s_t) =  \displaystyle\sum_{e_t = 0}^{M}p(E_t = e_t) \cdot p(S_t = s_t|E_t = e_t) 
\end{equation}

%\noindent $I_t$ -- Incidence at time $t$: 
The total number of infections $I_t$ at each point in time is $I_t = N_t + S_t$ and by the addition of two Poisson distributions that are assumed to be independent, i.e \autoref{eq: sseb_nt} and \autoref{eq:sseb_st} the incidence data of the SSEB model is defined to be: 
\begin{equation}
I_t | (E_t = e_t) \sim \text{Poisson}(\alpha R_0 \lambda_t + \beta e_t)
\label{eq:sseb_it}
\end{equation}

%\noindent \textbf{The Offspring Distribution}: 
In the SSEB model, the offspring distribution of a single individual encompasses the contribution due to non super-spreading events and the contribution due to super-spreading events as follows:
\begin{equation} 
Z \sim \text{Poisson}(\alpha R_0) + \text{Poisson}\left(\beta\left(\text{Poisson}\left(\dfrac{R_0(1 - \alpha)}{\beta}\right)\right)\right) 
\label{eq:z_sseb}
\end{equation}

The second term is a compound Poisson distribution. We can check that the expectation of $Z$ is $\mathbb{E}[Z] = \alpha R_0 + \beta \cdot \frac{R_0(1 - \alpha)}{\beta} = \alpha R_0 + R_0(1 - \alpha) = R_0$.

%% file: SECTION_MODELS/5_SSIB_MODEL.tex
The SSIB model is a bimodal model for super-spreading individuals. The model features two classes of infected individuals: non-super-spreading individuals \(N\) and super-spreading individuals \(S\), the latter being \(b\) times more infectious. Both classes can produce offspring that are either super-spreading individuals or not. The incidence data $I_t$ at time $t$ comprises non super-spreading individuals (infections), $N_t$ and super-spreading individuals (infections) $S_t$.  %The model parameters are the reproduction number \(R_0\), \(a\), and \(b\).
%
%\begin{itemize}
The parameter \(a\) represents the proportion of \(R_0\) attributed to non-super-spreading individuals, analogous to \(\alpha\) in the SSEB model. %Thus, \(1 - a\) corresponds to super-spreading contributions. The model reduces to the Baseline model when \(a = 1\).
The parameter \(b\) denotes the increased infectiousness of super-spreading individuals. Their infectivity profile is increased by a factor \(b\), 
    which is incorporated into the infectious pressure using the updated \(\lambda'_t\). The parameter \(\lambda'_t\) depends on the history of non super-spreading infections and super-spreading infections as follows:
\begin{equation}
\lambda_{t}^{'} \mid (\boldsymbol{N}_{[1:t-1]}, \boldsymbol{S}_{[1:t-1]}) = \sum_{\tau=1}^{t-1} \bigg( N_{\tau} + b \cdot S_{\tau} \bigg) \cdot \omega(t-\tau) \hspace{4mm} %\text{ for } t = 2, \dots, T  
\label{eq:lambda_dash}
\end{equation}

%\end{itemize}

%\noindent \textbf{The Incidence Model}: 
The incidence data $I_t=N_t+S_t$ comprises $N_t$ non super-spreading individuals and $S_t$ super-spreading individuals distributed as:
%
%\noindent \emph{Non-Super-Spreading Infections}: 
\begin{equation}
N_t \sim \text{Poisson}(a \cdot R_0 \cdot \lambda_t^{'}) % \hspace{8mm} \text{ for } t = 2, \dots, T$
\end{equation}
%
%\noindent \emph{Super-Spreading Infections}: 
\begin{equation}
S_t \sim \text{Poisson}\left(\frac{R_0(1 - a)}{b} \cdot \lambda_t^{'} \right) %\hspace{8mm} \text{ for } t = 2, \dots, T
\end{equation}
%\noindent $I_t$ -- \emph{Total Infections at time} $t$: 
%The incidence data $I_t$ at time $t$ comprises non super-spreading individuals, $N_t$ and super-spreading individuals $S_t$. 
%Individuals are assumed to infect susceptibles in the population independent of one other. The infections $N_t$ and $S_t$ are conditionally independent given the histories $\boldsymbol{N}_{[1:t-1]}$ and $\boldsymbol{S}_{[1:t-1]}$. 
%The incidence data $I_t$ is therefore the sum $I_t = N_t + S_t$ as follows:
%\begin{align}
%I_t \mid \boldsymbol{N}_{[1:t-1]}, \boldsymbol{S}_{[1:t-1]} & \sim \text{Poisson}\left(a \cdot R_0 \cdot \lambda_t^{'}\right) + \text{Poisson}\left(\frac{R_0(1 - a)}{b} \cdot \lambda_t^{'}\right)
%\end{align}

%\noindent \textbf{The Offspring Distribution}: 
To compute $Z$ we need to account for the contribution of transmission from both $N$ and $S$.
%
%\noindent \emph{Contribution to Z due to N}: 
For a non-super-spreading individual $N$, their offspring can be either non super-spreading individuals, contributing $Z_{N|N}$ to $Z$, or super-spreading individuals, contributing $Z_{S|N}$ to $Z$: 
\begin{equation}
Z_{N|N} + Z_{S|N} \sim \text{Poisson}(aR_0) + \text{Poisson}\left(\frac{(1-a)R_0}{b}\right) = \text{Poisson}\left(aR_0 + \frac{(1-a)R_0}{b}\right)
\end{equation}
%
%\noindent \emph{Contribution to Z due to S}: 
For a super-spreading individual $S$, their offspring can either be non super-spreading individuals, contributing $Z_{N|S}$ to $Z$, or super-spreading individuals, contributing $Z_{S|S}$ to $Z$: 
\begin{equation}
Z_{N|S} + Z_{S|S} \sim \text{Poisson}(abR_0) + \text{Poisson}((1-a)R_0) 
= \text{Poisson}\left(abR_0 + (1-a)R_0\right)
\end{equation}
The probability $p$ of being a super-spreading individual is $p = \frac{\frac{1-a}{b}R_0}{\frac{1-a}{b}R_0 + aR_0} = \frac{1-a}{1-a+ab}$. The total offspring distribution \(Z\) is a mixture of the two distributions with weights $1-p$ and $p$ respectively:
\begin{equation}
Z \sim (1-p) \text{Poisson}\left(aR_0 + \frac{(1-a)R_0}{b}\right) \bigoplus p \text{Poisson}\left(abR_0 + (1-a)R_0\right) 
\label{eq:z_ssib}
\end{equation}

We can check that the expectation of \(Z\) is $\mathbb{E}[Z] = (1 - p)\left(aR_0 + \frac{(1 - a)R_0}{b}\right) + p\left(abR_0 + (1 - a)R_0\right) = R_0.$

%% file: 3_INFERENCE.tex
We use a Bayesian framework for inference of model parameters. We sample from the posterior distributions of the model parameters using Markov Chain Monte-Carlo (MCMC). Consider a model $M$ with parameters $\boldsymbol{\theta} = (\theta_1, \theta_2, ..., \theta_p)$ and incidence data $\boldsymbol{I}_{[1:T]} = [I_1, I_2, \dots, I_T]$. In Bayesian inference, prior information $p(\boldsymbol{\theta}|M)$  about the parameters $\boldsymbol{\theta}$ is combined with information from the sample data contained within the likelihood $p(\boldsymbol{I}_{[1:T]} | \boldsymbol{\theta},M)$. The result of updating the prior distribution based on observed data is called the posterior distribution $ p(\boldsymbol{\theta} | \boldsymbol{I}_{[1:T]}, M)$. The posterior distribution contains all the available information about $\boldsymbol{\theta}$, is used to make estimates or inferences and allows us to quantify the uncertainty associated with our parameter estimates. In selecting our priors, we primarily opt for weakly informative priors to guide our analysis. These priors allow a balance between existing knowledge and the data-driven inference process. Our framework remains disease-agnostic, allowing for the selection of more informative priors tailored to specific datasets as required. The prior distributions chosen for our models are displayed in \autoref{tab:table_priors}. For the parameter $R_0$, common to all five models, an Exponential(1) distribution is used as the prior distribution throughout the study. This distribution with a mean centered on $1$ implies that a priori, we are not specifying whether the outbreak is likely to propagate throughout the population, $R_0 > 1$, or die out, $R_0 < 1$. %A further explanation of the priors chosen is provided in the Supplementary Material. 
The MCMC algorithms used for inference of the five models are all variations of the Metropolis Hastings algorithm. %and are also outlined in the Supplementary material. 
Data augmentation is required to evaluate the likelihoods of the SSI and SSIB models which is necessary to carry out inference of the models. 

\begin{table}[t]
\centering 
\fontsize{12}{14}
\begin{tabular}{|l|l|l|l|l|} %p{24mm}|}
\hline
  Models& Parameter & 
Prior Distribution &
 Prior support &
Prior Mean 
\\
\hline
{All} & ${R_0}$ & Exponential(1) & [0, $\infty$] & 1 \\ %& [0.025, 3.7] \\
\hline
 {SSE, SSI} & ${k}$ & Exponential(5) & [0, $\infty$] & 0.2 \\ %[0.05, 0.7] \\
\hline  {SSEB, SSIB} & $a$, ${\alpha}$ & Beta(2,2) & [0, 1] & 0.5 \\ %& [0.035, 0.965] \\
\hline {SSEB, SSIB} & ${b}$, ${\beta}$ & 1 + Gamma(3,3) & [1, $\infty$] & 10 \\ %& [2.9, 22.7] \\
\hline
\end{tabular}
%\captionsetup{font=small} % Adjust caption font size
\caption{The model parameters, their chosen prior distributions and the relevant metrics of the chosen distributions including the support and mean.}% and 95 \% quantiles.}
\label{tab:table_priors}
\end{table}

%% file: 4_MODEL_COMPARISON.tex
Model comparison is the process of comparing a set of candidate models, given data \citep{linhart1986model}.  
In Bayesian statistics this is achieved by computing Bayes factors which is a ratio of model evidence \citep{kass1995bayes, hoeting1999bayesian}. 
%Despite their power, methods in Category (2) are often overlooked due to challenges in computing the marginal likelihood. We perform multi-model comparison of our five models using the approach in (2), estimating each model's model evidence and computing posterior model probabilities. 
Given model $M$, data $\boldsymbol{I}_{[1:T]}$, parameter vector $\boldsymbol{\theta}$,
parameter prior $p(\boldsymbol{\theta} | M)$ and likelihood $p(\boldsymbol{I}_{[1:T]} | \boldsymbol{\theta}, M)$, the model evidence is  $p(\boldsymbol{I}_{[1:T]}  |  M) = \int p(\boldsymbol{I}_{[1:T]} | \boldsymbol{\theta}, M) p(\boldsymbol{\theta} | M) \:d \boldsymbol{\theta}$. Posterior model probabilities, \( p(M_i | \boldsymbol{I}_{[1:T]}) \), provide a comprehensive comparison of multiple models by normalizing their Bayes factors \citep{ando2010bayesian} as defined in \autoref{eq:post_probs}. In this equation $p(M_i)$ represents the prior probability on model \( M_i \). If no prior preference exists, the model prior is often uniform over all $m$ models; \( p(M_j) = 1/m \). Using this prior, selecting the most likely model reduces to choosing the one with the highest evidence.
\begin{equation}
p(M_i | \boldsymbol{I}_{[1:T]}) = \frac{p(\boldsymbol{I}_{[1:T]} | M_i) p(M_i)}{\sum_{j=1}^{m} p(\boldsymbol{I}_{[1:T]} | M_j) p(M_j)}
\label{eq:post_probs}
\end{equation}

To estimate the model evidence, we use a method that combines importance sampling with MCMC, as proposed by \citet{touloupou2018efficient}. This approach has been shown to be effective, especially in scenarios with missing data \citep{hudson2023importance, mckinley2020efficient}. Importance Sampling (IS) can be used to estimate properties of a target distribution by generating weighted samples from a proposal distribution that is generally easier to sample from. Given data $\boldsymbol{x}$ and proposal distribution $q$ the marginal likelihood can be written as:
\begin{equation}
    p(\boldsymbol{x} ) = \int_{\boldsymbol{\Theta}} p(\boldsymbol{x}  | \boldsymbol{\theta}) \frac{p(\boldsymbol{\theta})}{q(\boldsymbol{\theta})} q(\boldsymbol{\theta}) \, d\boldsymbol{\theta}
\end{equation}

An unbiased estimator of the model evidence is therefore:
\begin{equation}
\widehat{p}(\boldsymbol{x} ) = \frac{1}{N} \sum_{i=1}^{N} p (\boldsymbol{x}  | \boldsymbol{\theta}_i ) \frac{p(\boldsymbol{\theta}_i)}{q(\boldsymbol{\theta}_i)} \label{eq:phat}
\end{equation}

The effectiveness of the estimator depends upon the variance of the sampling estimate as outlined in \citet{touloupou2018efficient}. To minimize the variance, we want the proposal $q(\boldsymbol{\theta})$ to resemble the posterior as closely as possible. The proposal distribution \( q(\boldsymbol{\theta}) \) is typically derived from the MCMC output. \cite{clyde2007current} suggests using a multi-variate \emph{t}-distribution with the location and scale parameters estimated from the MCMC output. To ensure the proposal is over-dispersed relative to the target distribution we use a ``defense mixture", which reduces variance \citep{hesterberg1995weighted}. The mixing proportion $p$ in the defense mixture is typically set to 
0.95, ensuring the ratio of prior to proposal density remains bounded above by $1/(1-p)$ \cite{hesterberg1995weighted}. For priors, we use weakly informative Exponential(1) priors for model parameters, similar to \citet{hudson2023importance}. %, and perform sensitivity analysis to confirm robustness as detailed in Reference thesis. 
The defense mixture is:
\begin{equation}
q_\mathrm{D}(\boldsymbol{\theta}) = p \cdot q(\boldsymbol{\theta}) + (1 - p) \cdot p(\boldsymbol{\theta})
\end{equation}

%\noindent \emph{Importance Sampling Algorithm}: 
To obtain an estimate of the model evidence, $\widehat{p}(\boldsymbol{I}_{[1:T]})$ for each of the five models we use the following steps based on \cite{touloupou2018efficient}. (1) Obtain samples $\boldsymbol{\theta}$ from the posterior distribution $p(\boldsymbol{\theta} | \boldsymbol{I}_{[1:T]}, M_i)$ by fitting model $M_i$ to incidence data $\boldsymbol{I}_{[1:T]}$ using MCMC. (2) Derive the proposal distribution $q(\cdot)$, a parametric approximation of the posterior distribution, using the sample $\boldsymbol{\theta}$. For the uni-variate Baseline model, a student's t-distribution with three degrees of freedom is used. For the four multivariate models (SSE, SSI, SSEB, SSIB) a multivariate t-distribution (with three degrees of freedom) is used as the proposal distribution. For the SSIB model the proposal distribution is more involved. This is due to the augmented data; the super-spreading infections data $\boldsymbol{S}_{[1:T]}$ required for evaluation of the model's likelihood. For the model's continuous parameters ($R_0, a, b$) the multivariate t-distribution is used as above. For $\boldsymbol{S}_{[1:T]}$, a Dirichlet multinomial distribution is chosen as the proposal distribution. This distribution is often used to model categorical data. 
%The details of which are outlined in \textbf{Supplementary details} 
(3) Estimate the model evidence $\widehat{p}(\boldsymbol{I}_{[1:T]})$ as in \autoref{eq:phat}. 
(4) Repeat steps (1)-(3) to estimate the model evidence for all five models in the framework; $\widehat{p}(\boldsymbol{I}_{[1:T]} | M_{\text{Baseline}}), \widehat{p}(\boldsymbol{I}_{[1:T]} | M_{\text{SSE}}), \widehat{p}(\boldsymbol{I}_{[1:T]} | M_{\text{SSI}}), \widehat{p}(\boldsymbol{I}_{[1:T]} | M_{\text{SSEB}}) \text{ and } \widehat{p}(\boldsymbol{I}_{[1:T]} | M_{\text{SSIB}})$. 
(5) To carry out model comparison, the posterior model probabilities are calculated for all five models using \autoref{eq:post_probs}. The model with the highest posterior probability is selected as the best-fitting model.

%% file: 5_REAL_DATA_APPLICATIONS.tex
\vspace*{-0.5cm}
\subsection*{Parameter Inference on Simulated Data}

\begin{figure}[!t]
\centering
\includegraphics[height=0.56\textwidth]{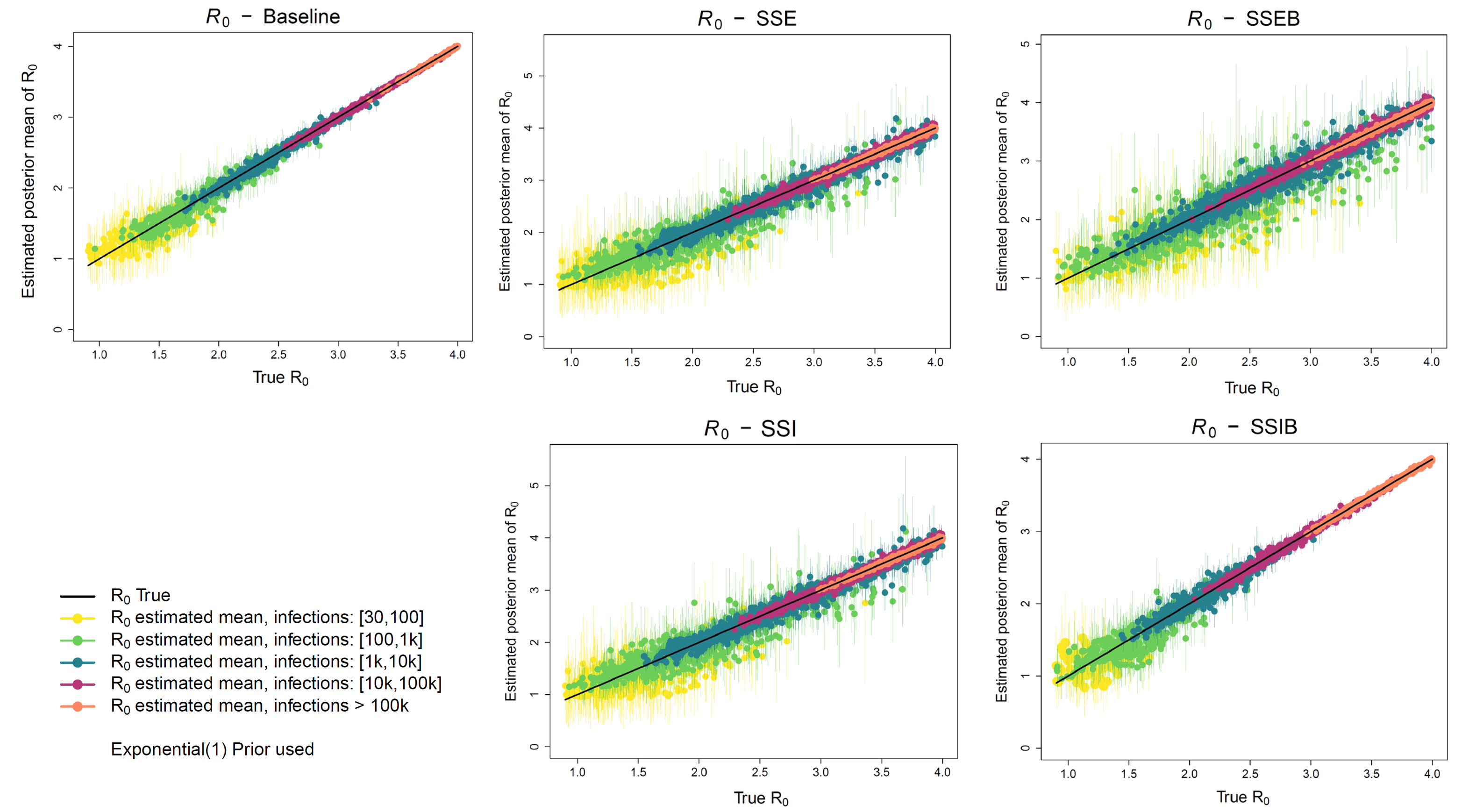}
\caption{Results of inference of $R_0$ across the five models. Posterior estimates %- mean \& 95 \% CI - of $R_0$ 
in each model
fit to 5000 simulated datasets from the model itself for $R_0$ in the range [0.9, 4.0]. Dots indicate the mean
of the estimated posteriors, bars represent the 95\% CI of the posterior, and the black diagonal line represents the true value used for simulation. 
%The mean and 95 \% CI of the Exponential(1) prior used in Bayesian inference is plotted in grey on the left. 
%In general inference is working well across the five models with the Baseline model and SSIB exhibiting the highest precision.
}
\label{fig:varied_results}
\end{figure}

An extensive simulation study was carried out to assess inference performance. Simulation studies provide
the unique advantage of knowing the underlying values used for generating the data, which can be used for comparison and validation of the estimates \citep{geweke2004getting}. The results of one such simulation study are displayed in \autoref{fig:varied_results}. The accuracy and precision of inference of the five models  (Baseline, SSE, SSI, SSEB, SSIB) in inferring $R_0$ across a range of relevant values are assessed allowing for a robust evaluation of inference performance across varying epidemic scenarios and data conditions. The results are based on 5000 simulated datasets generated from the models themselves with $R_0$ simulated across the range  $[0.9 - 4.0]$. In general inference is working well across the five models.
The total infection count of each simulated epidemic has an effect on the inferred results, as indicated by the color-coded simulations. Higher infection counts
(magenta, orange) have smaller biases and tighter 95\% CI compared to lower counts (yellow, green). 

\autoref{tab:fullsimu} contains the bias and coverage metrics for the inference of each parameter of each model.
The parameters used in the simulations were drawn uniformly from the range indicated. 
These results represent the mean values across 5000 independent repetitions of MCMC inference
applied to 5000 different simulations from each model. The simulations are of duration 50 days.
The bias (ie the mean difference between the parameter used in simulation and posterior mean) is small for all parameters of all models.
It is a bit higher for the $\beta$ parameter of SSEB and the $b$ parameter of SSIB, mostly because the natural range of these parameters is wider.
The coverage (ie probability that the 95\% CI contains the correct value used in the simulation) is around 95\% for all parameters of all models,
as expected under ideal conditions when the simulation and inference models are the same.

\begin{table}[H]
    \centering
    \begin{tabular}{|l|l|l|l|l|l|l|}
    \hline
        Model & Parameter & Range Tested & Bias & 95\% Coverage
         %(Mean) & MCMC Acceptance rate (\%) (Mean) & Effective Sample Size (Mean) 
         \\ \hline
        Baseline & $R_0$ & [0.9, 4.0] & -0.007 & 94 %& 39 & 4940 
        \\ \hline
        SSE & $R_0$ & [0.9, 4.0] & -0.017 & 94 %& 34 & 3530 
        \\ \hline
        ~ & $k$ & [0.01, 0.2] & 0.01 & 96 %& 35 & 4570 
        \\ \hline
        SSI & $R_0$ & [0.9, 4.0] & -0.016 & 94 %& 32 & 1785 
        \\ \hline
        ~ & $k$ & [0.05, 0.2] & 0.09 & 84 %& 32 & 420 
        \\ \hline
        SSEB & $R_0$ & [0.9, 4.0] & -0.02 & 95 %& 30 & 4200 
        \\ \hline
        ~ & $\alpha$ & [0, 1] & 0.0003 & 95 %& 30 & 4000 
        \\ \hline
        ~ & $\beta$ & [5, 15] & -0.16 & 97 %& 30 & 4500 
        \\ \hline
%        SSIB, known SS data & $R_0$ & [0.9, 4.0] & -0.005 & 95 %& 37 & 1000 
%        \\ \hline
%        ~ & $a$ & [0, 1] & 0.02 & 97 %& 37 & 370 
%        \\ \hline
%        ~ & $b$ & [5, 15] & 0.34 & 97 %& 37 & 400 
%        \\ \hline
        SSIB & $R_0$ & [0.9, 4.0] & -0.009 & 93 %& 28 & 1000 
        \\ \hline
        ~ & $a$ & [0, 1] & -0.05 & 92% & 28 & 500 
        \\ \hline
        ~ & $b$ & [5, 15] & 0.35 & 96 %& 28 & 300 
        \\ \hline
    \end{tabular}
    \caption{Performance metrics from the inference of parameters across the five epidemic
models. }
    \label{tab:fullsimu}
\end{table}

\subsection*{Model Comparison on Simulated Data}

To test our model comparison methodology we perform a simulation study using data generated from the five models before considering applications to real data. The goal is to determine how consistently model selection can select the correct simulation model as the most likely candidate among the five. For each analysis, we select one model as the simulation model, simulate 100 incidence datasets, and fit all five models to compute posterior model probabilities using \autoref{eq:post_probs}. For the prior model weight we put equal probability a priori on each model and so  $p(M_i) = 0.2$. The results of a simulation study are summarised in \autoref{tab: tab_mod_comp_heat}. The table contains posterior model probabilities (mean, CI) of the simulation study for all combinations of simulation model and fitted model. 
Consistently, the model with the highest posterior probability is the simulation model, aligning with our goal. For $n = 100$ simulations from the Baseline model, the Baseline model was selected 99\% of the time, with a mean posterior probability of 0.94 (CI: [0.77, 0.99]). Similarly, the SSE model was identified 93\% of the time (mean: 0.90, CI: [0.43, 1.0]), the SSI model 67\% of the time (mean: 0.63, CI: [0.06, 1.0]), the SSEB model 97\% of the time (mean: 0.896, CI: [0.68, 1.0]), and the SSIB model 63\% of the time (mean: 0.554, CI: [0.05, 0.99]). 

These results indicate strong performance in identifying the correct model, especially for the Baseline, SSE, and SSEB models, with some variability observed for the SSI and SSIB models, reflecting greater uncertainty. An additional result is that the super-spreading events models identify each other as the second most probable models, as do the super-spreading individuals models. These findings support our strategy of developing separate models of super-spreading events and super-spreading individuals and treating the two mechanisms as separate modes of super-spreading in relation to epidemic outbreaks. We also perform a prior sensitivity analysis, repeating the model comparison with both the original priors and uniform priors, across various scenarios, including different $R_0$ values and epidemic durations. %Detailed results are available in Reference thesis.

\input{TABLES/table_heat_map_model_comparison}

\newpage

\input{SECTION_REAL_DATA_APPLICATIONS/2_REAL_DATA_SARS_CANADA}

\input{SECTION_REAL_DATA_APPLICATIONS/1_REAL_DATA_COVID_NZ}

%% file: TABLES/table_heat_map_model_comparison.tex
\newcommand{\applycolorCI}[2]{%
    \ifdim #1 pt > 0.8 pt
        \cellcolor{red!70}#1 {#2}%
    \else\ifdim #1 pt > 0.6 pt
        \cellcolor{red!50}#1 {#2}%
    \else\ifdim #1 pt > 0.4 pt
        \cellcolor{orange!50}#1 {#2}%
    \else\ifdim #1 pt > 0.2 pt
        \cellcolor{yellow!50}#1 {#2}%
    \else
        \cellcolor{yellow!30}#1 {#2}%
    \fi\fi\fi\fi
}

%\newgeometry{margin = 1.5cm}
\begin{table}[!t]
\centering 
\fontsize{10}{12}
\begin{tabular}{|p{20mm}| p{18mm}|p{19mm}|p{18mm}|p{18mm}|p{18mm}|p{19mm}|}
%\hline
%& \multicolumn{6}{|c|}{\rule{0pt}{5ex} \bfcell{Posterior Probabilities of Fitted Models}  \rule[-2ex]{0pt}{0pt}} \\
\hline
& & \centering\arraybackslash{Baseline} &  \centering\arraybackslash{SSE} & \centering\arraybackslash{SSI} &
\centering\arraybackslash{SSEB} &
\centering\arraybackslash{SSIB} \\
\hline
\centering\arraybackslash{Simulation Model} & {Selection} \hspace{10mm} {Probability} & & & & &\\
\hline
Baseline & {99\%} & \textbf{\applycolorCI{0.94}{ \; \; [0.77, 0.99] }} & \applycolor{0.001} & \applycolor{0.024} & \applycolor{0.034} & \applycolor{0.001} \\
\hline 
SSE & {93\%} & \applycolor{0} & \textbf{\applycolorCI{0.90}{ \; \; [0.43, 1.0]}} & \applycolor{0} & \applycolor{0.10} & \applycolor{0}\\
\hline
SSI & {67\%} & \applycolor{0.09} & \applycolor{0.02} & \textbf{\applycolorCI{0.63}{ \; \; [0.06, 1.0]}} & \applycolor{0.10} & \applycolor{0.152}\\
\hline
SSEB & {97\%} & \applycolor{0.002} & \applycolor{0.101} & \applycolor{0} & \textbf{\applycolorCI{0.896}{ \; \; [0.68, 1.0]}} & \applycolor{0.001} \\
\hline
SSIB & {63\%} & \applycolor{0.19} & \applycolor{0} & \applycolor{0.232} & \applycolor{0.023} & \textbf{\applycolorCI{0.555}{ \; \; [0.05, 0.99]}} \\
\hline
\end{tabular}
\caption{Summary of estimated posterior model probabilities (mean and CI) for each combination of simulated model and fitted model for $N = 100$ simulations from each model in the left hand column. 
%The Baseline model is correctly selected as the most probable model 99\% of the time, the SSE model 93\% of the time, the SSI model 67\% of the time, the SSEB model 97\% of the time, and the SSIB model 63\% of the time. These results indicate high sensitivity for the Baseline, SSE, and SSEB models, with notable uncertainty for the SSI and SSIB models.
}
\label{tab: tab_mod_comp_heat}
\end{table}

%% file: SECTION_REAL_DATA_APPLICATIONS/2_REAL_DATA_SARS_CANADA.tex
\subsection*{SARS Outbreak, Canada 2003}

We now turn to the application of our modelling framework to real epidemic outbreaks. 
%We select specific time series of incidence data from major epidemics of the 21st century, namely the Covid-19 pandemic in New Zealand and the outbreak of SARS in Canada in 2003. 
We fit our models to the reported incidence datasets and infer the model parameters using our MCMC algorithms. 
%The outputs include estimates of the basic reproduction number ($R_0$) across the five models. 
We then apply our model comparison framework to determine the most likely model or `maximum a posteriori' of our five models when fit to the reported incidence data. 
Cases from the first time step are assumed to be imported cases, as in comparable methods \citep{cori2013new}. The infectivity profile is represented by the generation time distribution. % or serial interval distribution, depending on which data is available. %For our analysis of SARS-CoV-2 and SARS we consider the following studies; for SARS-CoV-2 \citet{li2020early} estimated a mean serial interval of 7.5 days, 95 \% CI (5.3, 19 days). \citet{Nishiura2020} examined early cases of Covid-19 in China and estimated the mean serial interval to be 4.7 days 95 \% CI (4.5, 4.9 days). We use a Gamma(6,1) distribution, with a mean of 6 and mode of 5, as it fits within the credible intervals of the serial interval estimates. 
For the outbreak of SARS in 2003 a serial interval distribution with a median of 6 days and an inter-quartile range of 4-9 days was previously reported \citep{lloyd2005superspreading}. We therefore use a discretised Gamma(6,1) as the generation time distribution.

\begin{figure}[!t] %[htb] [!t]
\centering
\includegraphics[width=15cm]{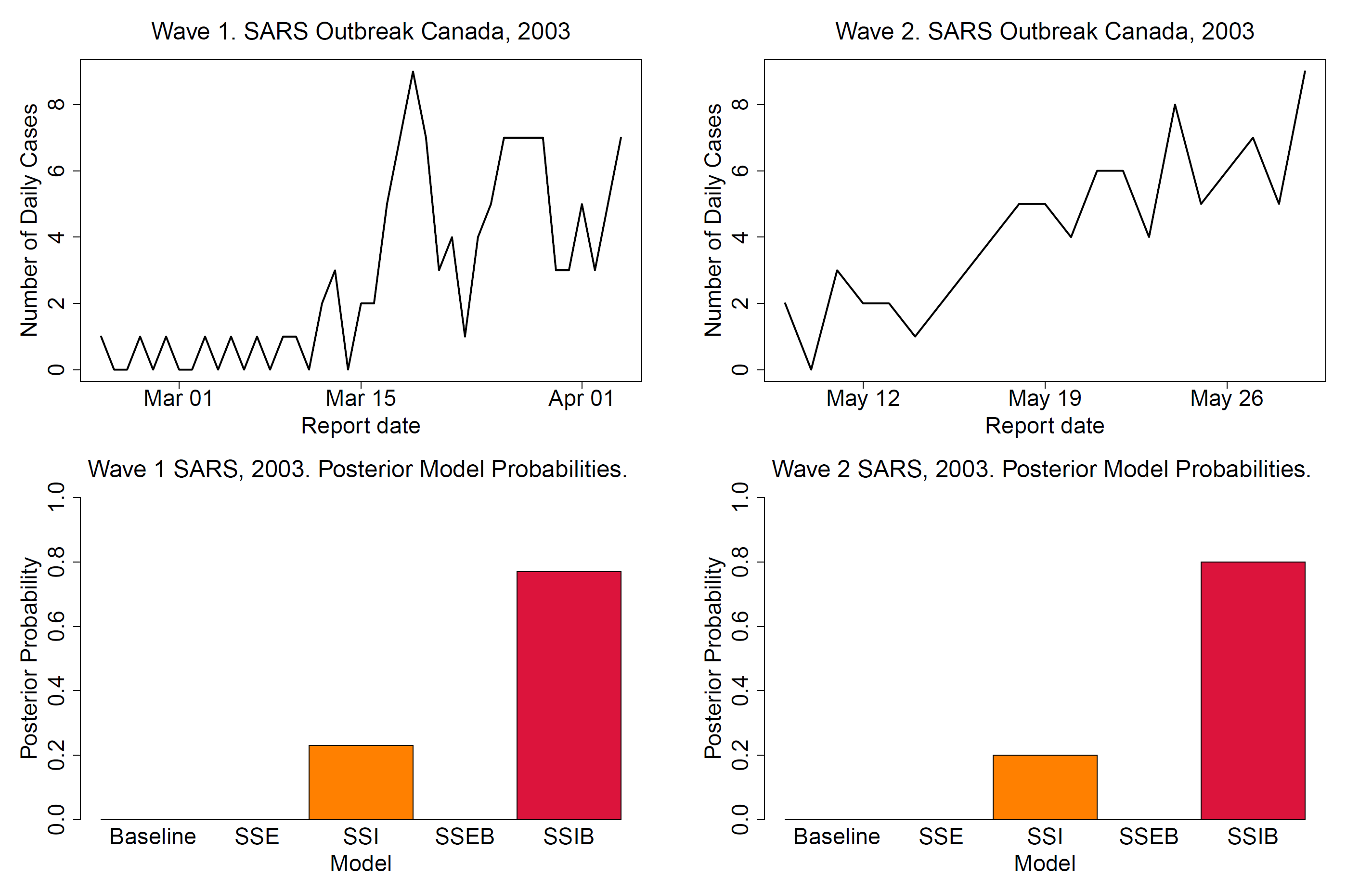}
\caption{Incidence data and bar plots of the posterior model probabilities of the five models applied to the two waves of SARS outbreaks in 2003.}
\label{fig:model_compare_sars}
\end{figure}

We apply our modelling framework to the outbreak of SARS in Canada in 2003. Severe Acute Respiratory Syndrome (SARS) is a viral respiratory illness caused by a coronavirus. The SARS outbreak originated in Asia in late 2002 and spread internationally. Canada's first case was reported in Toronto, Ontario, on February 23rd 2003 \citep{varia2003investigation}. The initial infection was traced back to a woman who had traveled to Hong Kong and returned to Toronto, resulting in a large nosocomial outbreak in a Toronto hospital, that yielded  128 infections \citep{varia2003investigation}. We speculate that this individual could be a super-spreading individual. For our analysis we focus on two waves of infection that see a spike in the number of reported cases in Canada (\autoref{fig:model_compare_sars}). 
%
%\subsubsection{Results}
%
%\textbf{Model Inference}: 
The $R_0$ estimates derived from MCMC are generally consistent across the five models, with low standard deviation observed for all models. For wave 1 the mean estimate of $R_0$ across the five models is 1.36 with standard deviation 0.04. For wave 2, the mean estimate of $R_0$ is 1.82 with standard deviation 0.96. 

%\noindent \textbf{Models Selection of SSI Models}: 
For both waves the super-spreading infections models emerge as the most likely models, specifically the SSIB model followed by the SSI model (\autoref{fig:model_compare_sars}). The SSIB model has a posterior model probability of 0.77 and 0.88 for the two waves respectively. The SSI model has the second highest posterior model probabilities of 0.23 and 0.20. The results align with reports in the literature that an individual caused a large nosocomial outbreak in a Toronto hospital, resulting in 128 infections \citep{varia2003investigation}. This suggests the presence of super-spreading individuals, as indicating by our model selection process. For the SSI model, $R_0 = 1.37$ (95 \% CI; 1.96, 2.0) for wave 1 and $R_0 = 1.75$ (95 \% CI; 1.2, 2.3) for wave 2, with dispersion parameter $k = 0.27$ (95 \% CI; 0.08, 0.51) for wave 1 and $k = 0.30$ (95 \% CI; 0.05, 0.62) for wave 2, indicating over-dispersion in secondary cases. For the SSIB model, $R_0 = 1.30$ (95 \% CI; 1.04, 2.0) for wave 1 and $R_0 = 1.85$ (95 \% CI; 1.3, 3.0) for wave 2. The parameter $a = 0.37$ (95 \% CI; 0.05, 0.76) for wave 1 and $a = 0.37$ (95 \% CI; 0.23, 0.53) for wave 2 further suggests that transmission is predominantly driven by super-spreading rather than homogeneous transmission.

%\newgeometry{margin=1.7cm}

%% file: SECTION_REAL_DATA_APPLICATIONS/1_REAL_DATA_COVID_NZ.tex
\subsection*{SARS-CoV-2 Outbreaks, New Zealand, 2020-2021}

We focus on outbreaks of SARS-CoV-2 in Aotearoa New Zealand in 2020 and 2021. In early 2020, New Zealand implemented a strict nationwide lockdown and closed its borders to all non-New Zealanders on March 20$^{\text{th}}$ 2020 \citep{cumming2022going} aiming for virus elimination \citep{geoghegan2020genomic}. As a geographically isolated island with borders that can be easily sealed, once closed, transmission could only occur at local or national level rather than from imported cases. The country also maintained a vigilant record of incidence cases throughout the pandemic and followed a four-level Alert Level framework \citep{cumming2022going, dpmc_covid19_group}. For these reasons we chose incidence data from New Zealand as one of the focuses of our analysis. 
For SARS-CoV-2 \citet{li2020early} estimated a mean serial interval of 7.5 days, 95\% CI (5.3, 19 days). \citet{Nishiura2020} examined early cases of Covid-19 in China and estimated the mean serial interval to be 4.7 days 95\% CI (4.5, 4.9 days). We use a Gamma(6,1) distribution, with a mean of 6 and mode of 5, as it fits within the credible intervals of the serial interval estimates. 

\begin{figure}[!t]
\centering
\includegraphics[width=15cm, height=6cm, keepaspectratio]{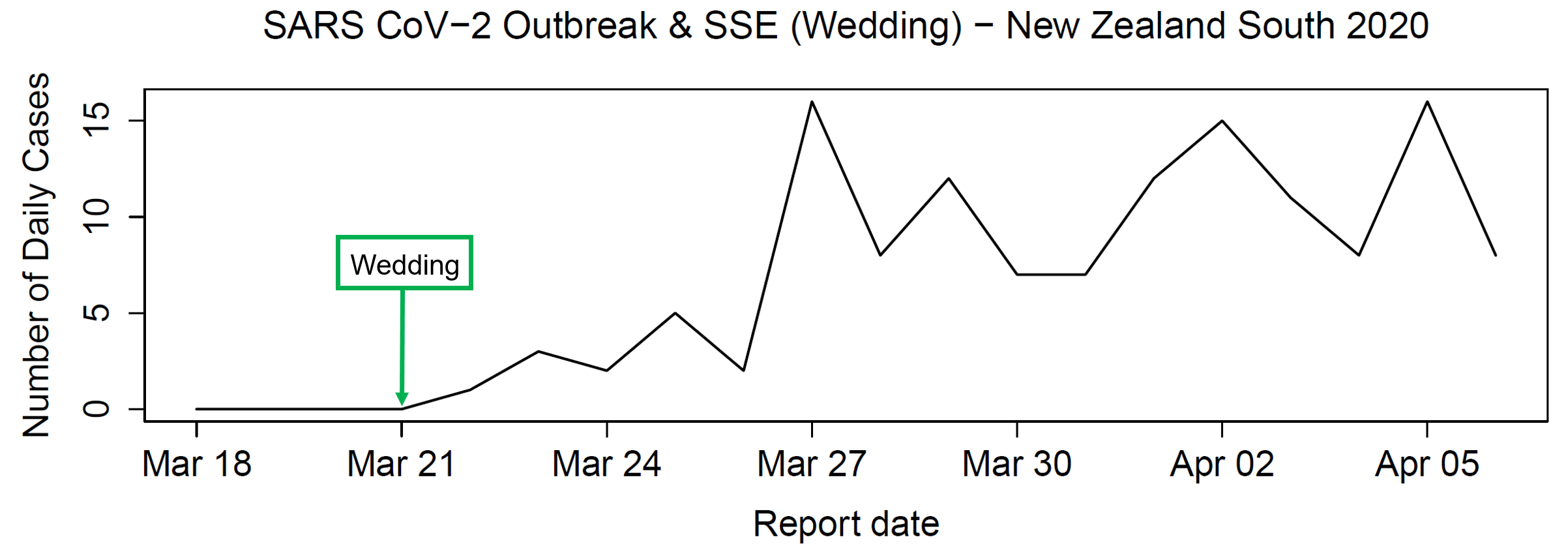}
\vspace{-0.5em} 
\caption{\small Incidence data from New Zealand's Southland district during the SARS-CoV-2 outbreak in 2020, before and after the wedding-related super-spreading event on March 21st. %\citep{minhealthnz_covid_data}
}
\label{fig:epi_wddding}
\end{figure}

%\noindent \textbf{Outbreak March 2020 - Wedding (Super-Spreading Event)}: 
In March 2020, a wedding in Bluff, Southland, led to New Zealand's largest cluster at the time, with 87 infections, and was classified as a super-spreading event \citep{nzherald2020bluff, stuff2020bluff}. We use case data from Southland before and after the event (\autoref{fig:epi_wddding}) \citep{minhealthnz_covid_data}. No cases were reported before March 21st, followed by a sharp rise after the wedding. Southland’s low population density (3.33 people per km$^2$) helped officials attribute the spike to this cluster rather than community transmission \citep{grant_southland_2008}. We apply our model framework to infer parameters and identify the most probable model, anticipating the SSE and SSEB models to be the best fit due to the super-spreading event. 

\begin{figure}[!t]
\includegraphics[width=16cm, height=6cm, keepaspectratio]{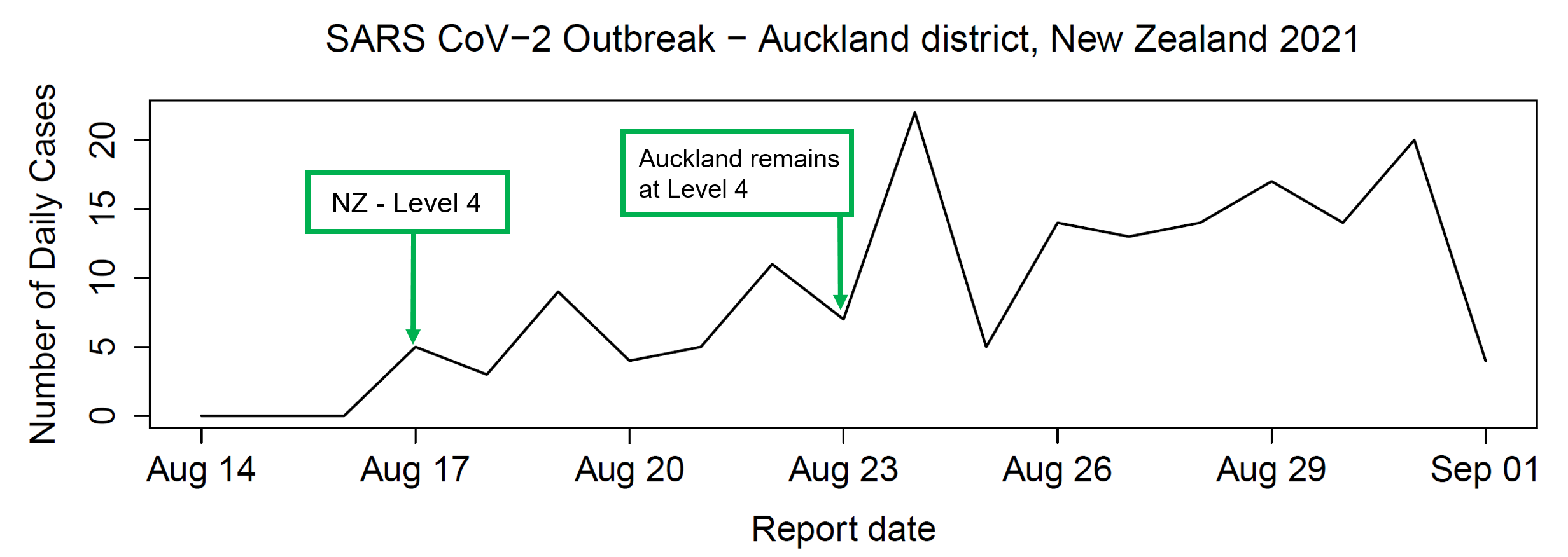}
\vspace{-0.5em} 
\caption{\small Reported cases from the Waitemata district of Auckland, New Zealand during a wave of SARS-CoV-2 cases in August 2021 resulting in a Level 4 lockdown. %\citep{minhealthnz_covid_data}
}
\label{fig: epi_waita}
\end{figure}

%\noindent \textbf{Outbreak August 2021 - Auckland}: 
We also analyze the August 2021 SARS-CoV-2 outbreak in Auckland, which coincided with New Zealand's Level 4 lockdown \citep{dpmc_covid19_group}. On August 17th, the government imposed the highest alert level nationwide, but while restrictions were eased elsewhere on August 23rd, Auckland remained under Level 4 \citep{cumming2022going}. We use case incidence data from Waitemata district of Auckland before and after this date (\autoref{fig: epi_waita}) and apply our model framework to infer parameters and identify the most probable model.

%\subsubsection{Results}

%\textbf{Model Inference}: 
We applied our model framework to the March 2020 and August 2021 New Zealand outbreaks, inferring parameters for the five models using MCMC. The basic reproduction number, $R_0$, estimates across models are shown in \autoref{tab:nz_r0} and are generally consistent. For the March 2020 outbreak, the mean $R_0$ is 2.11 with standard deviation 0.19, and for August 2021, the mean $R_0$ is 1.92 with standard deviation 0.13. For both outbreaks $R_0$ is significantly greater than 1.0, consistent with the fact that the outbreaks have not died out in the selected time series.

\input{TABLES/table_nz_r0}

\begin{figure}[!h] %[htb]
\centering
\includegraphics[width=12cm]{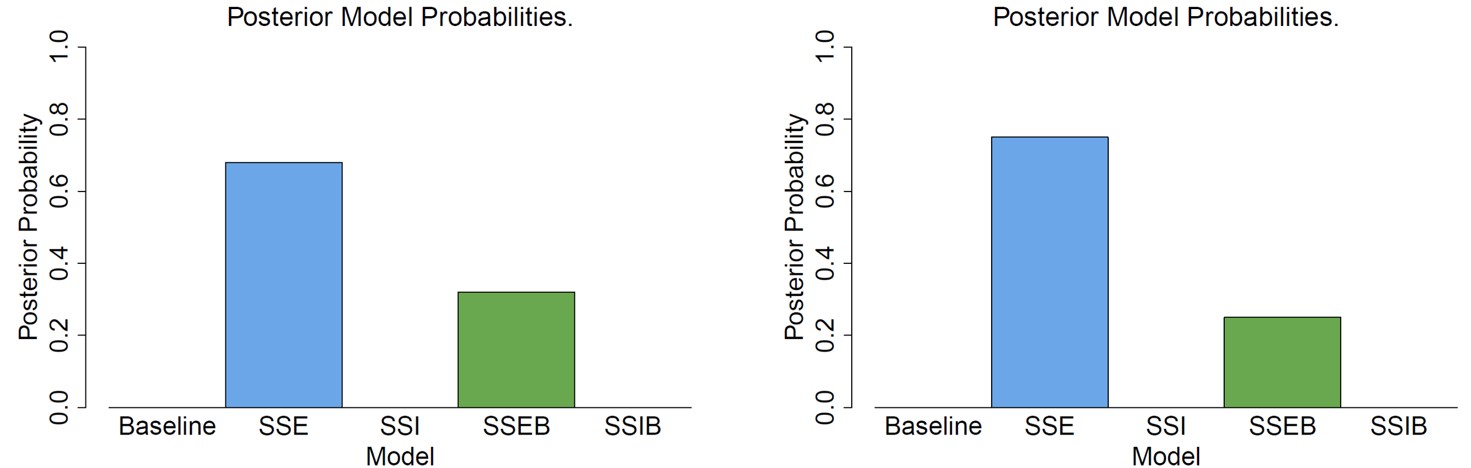}
\caption{Bar plots of the posterior model probabilities of the five models applied to the two waves of of SARS-CoV-2, New Zealand in 2020 and 2021.}
\label{fig:post_probs_nz}
\end{figure}

%MODEL COMPARISON
%\noindent \textbf{Model Comparison}: 
We performed model selection of our five models applied to the incidence data from New Zealand (\autoref{tab: post_probs_nz} and \autoref{fig:post_probs_nz}). Model evidence was computed using the importance sampling estimator from \autoref{eq:phat}, and posterior model probabilities were subsequently computed in order to carry out model selection.
%
%\noindent \textbf{Selection of Super-Spreading Event Models}: 
For both outbreaks, the SSE model is selected as the most likely, with posterior probabilities of 0.68 and 0.75, followed by the SSEB model at 0.32 and 0.25. These findings align with prior knowledge of super-spreading events in each outbreak; the Bluff wedding in 2020 and a church gathering in Auckland in 2021. 
\input{TABLES/table_posterior_probs_nz}

\input{TABLES/table_nz_best_fitting_parameters}

The parameter estimates for the best-fitting SSE and SSEB models are displayed in \autoref{tab:tab_nz_best_fit}.
For the SSE model, the 2020 Southland outbreak estimates are $R_0 = 2.26$, (95\% CI; $1.17, 3.37$) and $k = 0.32$ (95\% CI; $0.11, 0.57$). For the 2021 Auckland outbreak, $R_0 = 1.99$ (95\% CI; $1.3, 2.8)$ and $k = 0.41$ (95\% CI; $0.17, 0.7)$. The low estimates of $k$ indicate a high level of over-dispersion in the average number of secondary cases transmitted.  \citet{james2021model} estimated $k = 0.29$ (95\% CI; 0.10, 0.51) for SARS-CoV-2 in New Zealand using contact tracing data, which aligns with our estimates: $k = 0.32$ (95\% CI; 0.11, 0.57) and $k = 0.41$ (95\% CI; 0.17, 0.70). Notably, we are able to obtain comparable estimates of \(k\) from incidence data, similar to those obtained by fitting the offspring distribution to secondary case data. For the SSEB model's $\alpha$ parameter, estimates are $\alpha = 0.33$ (95\% CI; 0.02, 0.68) and $\alpha = 0.20$ (95\% CI; 0.05, 0.42). Since $1 - \alpha$ reflects the proportion of $R_0$ from super-spreading events, values of 0.67 and 0.80 suggest most transmission is due to super-spreading rather than homogeneous spread. The super-spreading parameter $\beta$ is estimated at 8.65 (95\% CI: 5.98, 11.61) for 2020 and $\beta = 4.67$,  95\% CI (2.85, 6.63) for the outbreak in 2021. Given that our novel bimodal model yields conclusions about super-spreading that align with those of the established SSE model strengthens the validity of our approach. This alignment underscores the potential of our models for use in a public health setting to identify the predominant modes of epidemic transmission.

%\begin{figure}[H] %[htb]
%\centering
%\includegraphics[width=14cm]{FIGURES/REAL_DATA/POST_PROB_PLOTS_NZ.PNG}
%\caption{Bar plots of the posterior model probabilities of the five models applied to outbreaks of SARS-CoV-2, New Zealand in 2020 and 2021. The SSE model followed by the SSEB model are selected as the most probable models.}
%\label{fig: post_probs_nz}
%\end{figure}

%% file: TABLES/table_nz_r0.tex
\begin{table}[!t]
\centering 
\fontsize{10}{12}
\begin{tabular}{|l|l|l|}
\hline
%& \multicolumn{2}{|c|}{\rule{0pt}{5ex} $R_0$ estimate -  Mean and 95 \% CIs  \rule[-2ex]{0pt}{0pt}} \\
%\hline 
Model &Outbreak, NZ South 2020 &  Outbreak, Auckland 2021  \\
\hline
Baseline & {2.30} [1.9, 2.7] & {2.0} [1.7, 2.3] \\
\hline 
SSE & {2.25} [1.18, 3.35] & {2.0} [1.3, 2.8]\\
\hline
SSI &  {1.80} [1.1, 2.9]&  {1.65} [0.85, 2.55]  \\
\hline
SSEB &  {2.10} [1.2, 2.9] & {1.90} [1.32, 2.55]  \\
\hline
SSIB & {2.10} [1.1, 3.08] & {2.0} [1.2, 3.26] \\
\hline
%\rowspace 
%Mean \& (sd) across models & 2.11 (0.19) & 1.92 (0.13) \\
%\hline
\end{tabular}
\caption{The mean and 95 \% CI of the estimates of $R_0$ across the five models when applied to incidence data from the SARS-CoV-2 Outbreaks in New Zealand in 2020 and 2021. 
%The mean and sd across the five models for each wave are also displayed.
}
\label{tab:nz_r0}
\end{table}

%% file: TABLES/table_posterior_probs_nz.tex
\begin{table}[!t]
\centering 
\begin{tabular}{|l|l|l|}
\hline
%& \multicolumn{2}{|c|}{\rule{0pt}{5ex}\makecell{\textbf{} \\ {Posterior Model Probabilities and Evidence} \\   \\ \textbf{}} \rule[-2ex]{0pt}{0pt}} \\
%\hline
%\rowspace 
\centering\arraybackslash{Model}& \centering\arraybackslash{Outbreak, NZ South 2020} &  \centering\arraybackslash{Outbreak, Auckland, NZ 2021} \\
\hline
Baseline & 0 (-210) & 0 (-110)\\
\hline  
SSE &  {0.68} (-56) & {0.75} (-66)  \\
\hline 
SSI & 0 (-126) & 0 (-75)\\
\hline
SSEB &  {0.32} (-57) & {0.25} (-67) \\
\hline 
SSIB & 0 (-85) & 0 (-140) \\
\hline
\end{tabular}
\caption{\small Posterior model probabilities of the models when fit to the reported cases of the specific regional outbreaks of SARS-CoV-2 in New Zealand in 2020 and 2021.}
\label{tab: post_probs_nz}
\end{table}

%% file: TABLES/table_nz_best_fitting_parameters.tex
\begin{table}[!t]
\centering 
\fontsize{10}{12}
\begin{tabular}{|l|l|l|l|}
\hline
Model & 
Parameter & 
Outbreak, NZ South 2020&
Outbreak, Auckland, NZ 2021
\\
\hline
%\specialrule{0.15em}{0.2em}{0.2em} % Add horizontal line
{SSE} & ${R_0}$ &  {2.25} [1.18, 3.35] & {2.0} [1.3, 2.8]  \\
\hline
& ${k}$ & {0.32} [0.11, 0.57] & {0.41} [0.17, 0.7] \\
\hline
{SSEB} & ${R_0}$ &  {2.10} [1.2, 2.9] & {1.9} [1.32, 2.55] \\
\hline
& ${\alpha}$ &  {0.33} [0.02, 0.68] & {0.2} [0.01, 0.42] \\
\hline
& ${\beta}$ &  {8.65} [5.98, 11.61] & {4.67} [2.85, 6.63] \\
\hline
\end{tabular}
\caption{The model parameter estimates (mean and 95\% CI) of the SSE and SSEB models - models selected with the highest posterior model probabilities -  when applied to incidence data from New Zealand in 2020 and 2021.}
\label{tab:tab_nz_best_fit}
\end{table}

%% file: 6_DISCUSSION.tex
%DISCUSSION
The primary aim of this work is to develop a modelling framework for incidence time-series data that encompasses distinct models for super-spreading events and super-spreading individuals. Additionally, it seeks to perform model selection between these candidate models when fit to both simulated and, more critically, real epidemic data. To achieve this a comprehensive framework of stochastic branching process models of epidemic transmission is developed for time-series of incidence data that encompasses five distinct models. The SSEB and SSIB models presented introduce novel bimodal approaches to characterize super-spreading events and individuals. The models include distinct mechanisms of epidemic transmission. 

We have successfully implemented MCMC algorithms to infer the parameters of all five models. The results of our quantitative inference study highlight the effectiveness of the inference methods. This is particularly true for the basic reproduction number $R_0$ across all models and is also supported when we apply our methods to incidence time-series of real outbreak data. Our estimates of \( k \) align closely with those obtained in studies using secondary case data, underscoring the utility of incidence data, which is more widely available. For example, while some studies rely on detailed contact tracing to estimate transmission dynamics \citep{james2021model}, our models achieve similar insights using temporal data alone. This capability could significantly streamline surveillance during outbreaks by reducing reliance on resource-intensive data collection. We can also infer the novel parameters of our novel super-spreading models, namely $\alpha$ and $\beta$ of the SSEB model and $a$ and $b$ in the SSIB model; when the super-spreading infections data is available. Interestingly when we compare the estimates of our best fitting SSE and SSEB models to the outbreak of SARS-CoV-2 in New Zealand for example, we observe low values of both $k$ and $\alpha$ in the SSE and SSEB models respectively. In the definition of our model this corresponds to a large amount of super-spreading or over-dispersion in both cases. Similarly for the outbreak of SARS in Canada in 2003 we estimate low values for both $k$ and $a$ in the best fitting SSI and SSIB models. Such results provide support for our SSEB and SSIB models as novel methods for quantifying super-spreading.

Model comparison is an important feature of our framework. Across simulations, the true simulation model consistently achieves the highest posterior probability, confirming that the five models are both identifiable and capable of capturing distinct mechanisms of epidemic transmission. When tested with real outbreak data, the model selection results exhibit clear and consistent trends. We extract time-series of distinct time periods of each disease and region. Yet the same models are consistently selected as the best fit for the same disease and region at different points in time, while different models are selected for different diseases. This suggests that the mechanisms of epidemic transmission of some of our models are a better fit to certain diseases than others. The SARS outbreak in 2003, which was controlled to an extent and did not escalate into a global pandemic, was best described by the SSI and SSIB models. In contrast, for SARS-CoV-2, which caused the major Covid-19 pandemic, the SSE and SSEB models, which account for super-spreading events, provided the best fit. This aligns with the observed epidemiology of SARS-CoV-2, where super-spreading events played a significant role in transmission dynamics \citep{lewis2021superspreading,du2022systematic,brainard2023super}. These models emphasize individual-level heterogeneity. Notably the Baseline Poisson model, that assumes homogeneous transmission in the population with no capacity for over-dispersion, i.e. equal mean and variance, is never selected as the best fitting model. This provides further support for the necessity of incorporating dispersion or super-spreading in models for epidemic transmission. 

%\emph{Limitations}: 
%We now address some of the limitations of the work presented in this thesis. 
From an epidemiological standpoint, the simplicity of the models, with a maximum of three parameters, is both a strength and a limitation. While the simplicity aids in the clarity and applicability of the models, the models may not capture the full extent of the complexities of real-world epidemics where super-spreading events and individuals often occur simultaneously. A more complex model combining both SSE and SSI could potentially provide a more accurate representation. However there are benefits to keeping models as simple as possible, such as their rapid implementation and flexibility across use-cases.
We only consider constant parameters, we do not consider variations of our parameters over time, for example a time-varying reproduction number $R_t$ \citep{cori2013new} or time-varying $k_t$ \citep{ho2023accounting, adam2022time}. This limitation means that our models may not capture the dynamic nature of epidemic spread, where parameters can change over time due to factors such as public health interventions, changes in population behavior, or pathogen evolution. However, we focus on relatively short time windows, which somewhat mitigates this limitation.
Additionally, the assumption of complete reporting of infectious cases overlooks potential sampling biases in real-world data. While this is a common assumption in the literature \citep{cori2013new}, it remains a potential source of inaccuracy. A potential solution to this would be incorporating the sampling proportion as a parameter in our models. However, considering a partially observed branching process can be complicated. It requires accounting for the possibility that entire sections of a transmission tree, which depicts the chain of transmission events within a population, may be observed or missed. This complexity arises because the available data might not capture every individual case or transmission event, making it challenging to accurately reconstruct the full transmission dynamics \citep{didelot2017genomic,carson2024inference}. %The availability of high-quality incidence data posed further constraints, as we were unable to access outbreak time-series data from key regions like Hong Kong and Singapore for the 2003 SARS epidemic. Similarly, limited access to contact tracing data, which is often not publicly available, restricted comparisons to dispersion parameter estimates derived from such datasets. Finally, the models assume an unlimited pool of susceptibles, excluding effects of vaccination or immunity. 

%\emph{Future work}. 
%There are numerous possible extensions to this work. Given the relevance of the offspring distribution in identifying super-spreading individuals, we could enhance our model fit by combining the incidence time series data with offspring distributions by multiplying their likelihoods. This could provide an even more complete picture of the transmission dynamics of the diseases. However this requires accurate observation of transmission chains, typically gathered through contact tracing or phylodynamic analysis, which can often be influenced by reporting bias, estimation methods, and transmission scenarios \citep{du2022systematic}. 
%
In conclusion, this work presents a novel modelling framework using time series incidence data to analyze super-spreading phenomena, offering an alternative to models that rely on less accessible secondary case data. We developed and validated five distinct stochastic branching-process models, alongside a robust comparison method that differentiates them, capturing unique epidemic transmission mechanisms. Our approach consistently identifies transmission mechanisms across different time series, highlighting the inadequacy of homogeneous transmission assumptions and emphasizing the need for models that incorporate dispersion and super-spreading. Quantifying the impact of super-spreading is crucial for disease control; focusing efforts on super-spreading events could significantly reduce the reproduction number, as previously noted \citep{endo2020estimating}. Public health initiatives can leverage our models to target super-spreading events and individuals for tailored interventions, as seen in Japan during the COVID-19 pandemic \citep{ueda2023identifying}. This targeted approach allows for efficient resource allocation, potentially curbing epidemic spread more effectively than generalized strategies. Overall, our modelling framework provides valuable insights for public health officials, enhancing epidemic management and prevention.